# Crossing Stocks and the Positive Grassmannian I: The Geometry behind Stock Market

Ovidiu Racorean

decontatorul@hotmail.com

**Abstract**

It seems to be very unlikely that all relevant information in the stock market could be fully encoded in a geometrical shape. Still, the present paper will reveal the geometry behind the stock market transactions. The prices of market index (DJIA) stock components are arranged in ascending order from the smallest one in the left to the highest in the right. In such arrangement, as stock prices changes due to daily market quotations, it could be noticed that the price of a certain stock get over /under the price of a neighbor stock. These stocks are crossing. Arranged this way, the diagram of successive stock crossings is nothing else than a *permutation diagram*. From this point on the financial and combinatorial concepts are netted together to build a bridge connecting the stock market to a beautiful geometrical object that will be called **stock market polytope**. The stock market polytope is associated with the remarkable structure of *positive Grassmannian* .This procedure makes all the relevant information about the stock market encoded in the geometrical shape of the stock market polytope more readable.

1. Introduction

A general stock market index is a method to measure the state of a market evaluating the prices of a selection of stocks owned by the most representative companies. These selected stocks are the components of the index. The market index is computed typically by a weighted average of the selected stocks prices and the performance of stock market can be asses.

Such a global view of market performances, although extremely valuable, is not offering an image of what happen with the stock components in the evolution of the market quotations. Viewing the relations between stock components of a market index on a daily basis gave valuable complementary information that help in the investment decision making process.

Prices of the stock index components can be arranged in a table in ascending order starting from smaller stock prices, in the left, to companies having higher stock prices, to the right. Section 2 is devoted to explain the concept in detail, here is suffice to underline the main reason for such stock order. It could be noticed that along the time some stocks surpass or come under the price of neighbors stocks in the parallel series. Other way to see this is to imagine the stocks are crossing their price series.

To exemplify the stocks crossing diagram with examples from the real stock market, the market index is chose to be *Dow Jones Industrial Average* (DJIA) with its components. Prices of a fraction of all 30 components of DJIA are arranged in the manner explained above, starting from CSCO which is the lowest priced stock until the highest, PG, for the market reference date 5/15/2013.

To simplify the understanding of these particular diagrams only 4 stocks are retained and the prices wiring diagram is drawn in section 3, along with relevant examples of stock crossings.

This simply technic of viewing the stocks prices of the DJIA components reveals the deep connection that exists between stocks crossings and permutation diagrams. It can be said that every time the price of a stock in an index go up or under its neighbor price that the two stocks permute.

A deeper insight into measuring the performances of a market by examining the permutations of the stock components of a representative index on that market can be possible by building a bridge that directly connects stock markets to more complex structures that exist in combinatorial mathematics, such as decorated permutations, hook diagrams or positroids.

Sections 4 and 5 will constitute an insight on the last years work of the mathematician Alex Postnikov[1],[2] related to connections between wiring diagrams , decorated permutation, and hooks. Examples on how these beautiful combinatorial objects are related to the states of the stock market are provided.

The last bricks in the construction of the bridge relating stock market to geometry behind it are shown in section 6. New combinatorial structures are coming into the "scene" and are explained in association with financial concepts, in their order of "apparition". The remarkable structures of matroids, positroids are effectively connected to price quotations of stocks in the market, such that a new geometrical object emerges, the ***stock market polytope***.

Polytopes are geometrical structures that live in higher dimensions such that is typically impossible to represent them graphically. The example of the chosen four DJIA components is exceptionally provided a beautiful polytope structure depicted at the end of this section.

Section 7 is a refinement of the combinatorial approach to stock market, intermediated by Grassmann geometry. The reader is briefly exposed to the more complex geometry of *positive Grassmannians* , intensively studied in the latter years, mainly due to many applications of it in Quantum Field Theory, see for details [3],…,[6]. Technical terms like *positive Grassmannian cells* and positroid decomposition may look complicated, but are easy to be explained as removals of some stock crossings, and help in deeper understanding the properties of stock market polytope.

Considering the DJIA example of a market index, it is straightforward to see that because of the large number of components, the stock market polytope associated to it will be complicated and that it won't be easy to compute. Still, using dedicated software like MATEMATHICA it should relax the process to reveal the shape of the usual stock market polytope or to define geometrical structure of market under stress, like the 2008 crisis.

The volume of the stock market polytope can be determined and will evaluate the probability of a certain state of stock market, but this will be the subject of a future papers.

2. **Briefing of a general stock market index**

The idea to measure the average performance of the stock market as a whole is old and can be traced back in the mid-1880 when Charles Dow, Edward Jones and Charles Bergstresser launched for the first time the concept of market index. Their first attempt to index the market contained nine stock components owned by the most influential companies of that time. In 1896 *Dow Jones Industrial Average* was born in its own rights.

A stock market index is a method to measure the performances of a market by a selection of the most representative stocks that are traded on the respective market. It is computed typically by a weighted average of the selected stocks prices. The levels of the market index indicate the state of the market, a market in contraction for small index levels or a market in expansion for high levels of the index.

Such a global view of market performances, although extremely valuable, is not offering a complete image of what happen with the stock components in the evolution of the market quotations, how the relations and correlation between them changes along the time. Some attempts to reveal the relations between market index stock components were made to cover particular periods in the market existence such as the 2007 financial crisis, see for details [13].

Although, viewing the relations between stock components of a market index on a daily basis is not crucial for market professionals, it gave valuable complementary information that help in the investment decision making process. The present paper constitutes a unique and completely different attempt to offer investors rapid images of the market moves. In the following sections a map relating stock market with complex geometrical objects is created, such that at the end of the paper any investor would be capable to interpret the market moves only by seeing geometrical shapes behind stock market transactions.

To simplify the discussion and give examples from real stock market conditions, and also, to celebrate its long existence, the index of interest is choosing to be the *Dow Jones Industrial Average* (DJIA). A fraction of price quotations for some DJIA components are shown in table 1 as daily closing prices for a period in 2013 between 5/15/2013 and 6/7/2013.

| Date | CSCO | GE | INTC | PFE | MSFT | T | KO | MRK | VZ | JPM | DD | DIS | NKE | UNH | AXP | HD | WMT | PG |
|------|------|------|------|------|------|------|------|------|------|------|------|------|------|------|------|------|------|------|
| 6/7/2013 | 24.5 | 23.9 | 24.6 | 28.3 | 35.7 | 35.5 | 41.4 | 48.2 | 50.2 | 54.3 | 55.4 | 64.9 | 62.8 | 62.6 | 78 | 78.7 | 76.3 | 77.8 |
| 6/6/2013 | 24.6 | 23.4 | 24.7 | 28.1 | 35 | 35.8 | 40.8 | 48.6 | 50 | 53.5 | 54.8 | 63.1 | 62.2 | 61.9 | 76.2 | 77.3 | 75.6 | 76.8 |
| 6/5/2013 | 24.3 | 23.3 | 24.7 | 27.5 | 34.8 | 35.3 | 40.7 | 48.7 | 48.3 | 53 | 54.6 | 63.1 | 61.8 | 61.8 | 74.8 | 75.1 | 75.3 | 76.7 |
| 6/4/2013 | 24.4 | 23.7 | 25.4 | 27.7 | 35 | 35.7 | 41.4 | 49.4 | 48.8 | 54 | 55.8 | 64.4 | 62.8 | 62.4 | 76.1 | 76.6 | 75.9 | 77.4 |
| 6/3/2013 | 24.4 | 23.6 | 25.2 | 27.8 | 35.6 | 35.1 | 40.8 | 48.5 | 48.7 | 54.5 | 56.1 | 63.8 | 63 | 62.8 | 76.5 | 79.1 | 75.7 | 77.7 |
| 5/31/2013 | 24.1 | 23.3 | 24.3 | 27.2 | 34.9 | 35 | 40 | 46.7 | 48.5 | 54.6 | 55.8 | 63.1 | 61.7 | 62.6 | 75.7 | 78.7 | 74.8 | 76.8 |
| 5/30/2013 | 24.4 | 23.6 | 24.2 | 28.3 | 35 | 35.5 | 40.8 | 47.1 | 49.1 | 55.6 | 56.3 | 64.7 | 62.4 | 64.7 | 76.1 | 79.4 | 75.6 | 79.1 |
| 5/29/2013 | 24.1 | 23.6 | 24.3 | 28.3 | 34.9 | 35.9 | 41.4 | 46.9 | 49.6 | 54.7 | 56 | 66.3 | 62.9 | 63.4 | 75.8 | 79.5 | 76.2 | 78.9 |
| 5/28/2013 | 23.9 | 23.6 | 24.1 | 29 | 35 | 36.2 | 42.6 | 47.6 | 50.8 | 54.6 | 55.9 | 66.7 | 63.3 | 63.3 | 76.2 | 79.8 | 77.3 | 80.9 |
| 5/24/2013 | 23.5 | 23.5 | 23.9 | 29 | 34.3 | 36.8 | 42.2 | 47.2 | 51.4 | 53.7 | 55.4 | 65.5 | 62.8 | 62.1 | 75.3 | 79 | 77.3 | 81.9 |
| 5/23/2013 | 23.5 | 23.7 | 24.1 | 29.1 | 34.2 | 36.7 | 41.9 | 47.3 | 51.9 | 53.4 | 55.4 | 65.2 | 63.3 | 62.4 | 74.7 | 78.9 | 76.3 | 78.7 |
| 5/22/2013 | 23.3 | 23.9 | 24.1 | 29.3 | 34.6 | 36.6 | 42.3 | 46.7 | 51.5 | 53.6 | 55.6 | 65.6 | 64.5 | 62.3 | 74.4 | 79.7 | 77 | 78.8 |
| 5/21/2013 | 24 | 23.7 | 24.2 | 28.8 | 34.9 | 36.9 | 42.3 | 47.3 | 52.1 | 53 | 56.4 | 65.8 | 65.2 | 62.9 | 75.1 | 78.7 | 77.4 | 78.8 |
| 5/20/2013 | 24 | 23.6 | 24.1 | 28.7 | 35.1 | 37.2 | 42.4 | 45.2 | 52.7 | 52.3 | 55.9 | 66.1 | 65.3 | 62.6 | 74.4 | 76.8 | 77.4 | 79.1 |
| 5/17/2013 | 24.2 | 23.5 | 24 | 29 | 34.9 | 37.4 | 43 | 46 | 53.4 | 52.3 | 55.9 | 66.6 | 65.3 | 62.8 | 73.3 | 76.9 | 77.9 | 80 |
| 5/16/2013 | 23.9 | 23.3 | 23.9 | 29.3 | 34.1 | 37.4 | 43.1 | 46.4 | 53.2 | 51 | 55.5 | 66.5 | 64.4 | 62.1 | 72.2 | 76.8 | 78.5 | 80.2 |
| 5/15/2013 | 21.2 | 23.2 | 24.2 | 29.6 | 33.9 | 37.5 | 42.9 | 46.7 | 53.6 | 51.1 | 55.6 | 67.7 | 65.8 | 61.6 | 72.8 | 77.9 | 79.9 | 80.7 |

**Table 1**. A fraction of the DJIA index components sorted by price quotations from the left to the right at 05/15/2013.

It can be easily seen that the DJIA stock components in the table 1 are arranged in ascending order from the stock with the smallest price quotation (CSCO) at the left to the stock with the highest price (PG) at the right at the start date 5/15/2013.This will be a rule of arranging the stock components of the DJIA index.

Notice that for others rows of the table this rule is not applied, such that next day, at 5/16/2013, for example, the price of CSCO came over the price of GE. The next section will explore the consequences of arranging all the rows in the table containing the stock components of DJIA in ascending order from the left to the right and some unexpected and beautiful combinatorial results will be revealed.

3. **Crossing stocks and permutations**

From this section on the financial and mathematical concepts, as combinatorial and geometrical aspects, are netted together step by step, from simply to complex, until the final geometrical shape picture of stock market is reached.

To simplify the exposition only four stock components of DJIA are retain further, AXP, HD, WMT and PG. The number of stocks is chosen such that the discussion should be neither trivial, nor too complex.

The price quotations for the chosen four DJIA components are arranged as it was stated in the first section, in ascending order from the stock with the smallest price (AXP) on the left to the stock having the highest price (PG) on the right at the starting date 5/15/2013. The arrangement of stocks from the left to the right in ascending order of prices is preserved for every row in the table, say for every trading day. In this manner the stocks prices will be shifted from their initial positions, at the right or left, every time the price of one of the four stocks comes under or over the price of the neighbor stock, put it in other words every time the ***stocks are crossing***. Crossings of stocks have a direct and important impact reflected in the value DJIA index, but this is a subject of a future paper. Here, just bear in mind intuitively that when stocks are crossing, the price of the stock that goes under its neighbor, the others remaining barely unchanged, will make the value of DJIA index to fall.

To have a clear visual image of the prices belonging to a certain stock, the time series of every stock is drown in a different color. This way of viewing the stock prices time series makes visible the crossing of stocks. Following the colors of stocks time series, the trajectories of price quotations can be interpreted as the figure 1 shown:

| | 5/15/2013 | 5/16/2013 | 5/17/2013 | 5/20/2013 | 5/21/2013 | 5/22/2013 | 5/23/2013 | 5/24/2013 | 5/28/2013 | 5/29/2013 | 5/30/2013 | 5/31/2013 | 6/3/2013 | 6/4/2013 | 6/5/2013 | 6/6/2013 | 6/7/2013 |
|---|---|---|---|---|---|---|---|---|---|---|---|---|---|---|---|---|---|
| AXP | 72.78 | 72.23 | 73.32 | 74.4 | 75.11 | 74.44 | 74.69 | 75.27 | 76.16 | 75.83 | 75.63 | 74.84 | 75.69 | 75.94 | 74.76 | 75.63 | 76.33 |
| HD | 77.88 | 76.75 | 76.86 | 76.76 | 77.39 | 77.03 | 76.33 | 77.31 | 77.32 | 76.23 | 76.14 | 75.71 | 76.47 | 76.06 | 75.25 | 76.24 | 77.75 |
| WMT | 79.86 | 78.5 | 77.87 | 77.4 | 78.71 | 78.82 | 78.7 | 78.99 | 79.82 | 78.9 | 79.09 | 76.76 | 77.66 | 76.63 | 75.1 | 76.82 | 78.04 |
| PG | 80.68 | 80.2 | 80.02 | 79.09 | 78.8 | 79.69 | 78.91 | 81.88 | 80.86 | 79.49 | 79.44 | 78.66 | 79.08 | 77.37 | 76.66 | 77.26 | 78.74 |

**Figure 1**. Crossings of stocks in fraction of 4 DJIA stock components.

Notice from the figure 1 above that from time to time ***stocks are crossing.*** As an example, it can be seen that at 5/21/2013 the closing prices of HD and WMT are crossing. As it was stated earlier the impact of the crossing on the total value of DJIA is, for now, neglected and the attention is focused only on the colored trajectories the stocks prices take in the DJIA components table.

Bearing in mind only the colored trajectories of stocks prices, totally neglecting the values of quotations, the figure 1 can be transpose in the picture bellow, where the crossings of stocks become very clear:

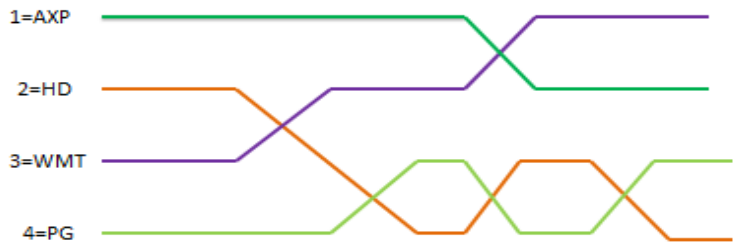

**Figure 2.** Wiring diagram for the crossing of stocks.

The resulting diagram is known in combinatorial mathematics as ***wiring diagram*** and is associated to permutations. A **permutation** is simply related to arranging and rearranging of a certain set of values, as is the case with the price of stocks that compose a market index.

The permutations are noted $\pi = \begin{pmatrix} 1 & 2 & \dots & n \\ \pi_1 & \pi_2 & \dots & \pi_n \end{pmatrix}$ or simply $\pi = \{\pi_1, \pi_2, \dots, \pi_n\}$ and states that a set of $n$ elements is arranged in $\pi_n$ way. Focusing the attention on the wiring diagram 3, the permutation associated to it is $\begin{pmatrix} 1234 \\ 2413 \end{pmatrix}$ or $\{2,4,1,3\}$, and simply says that 1 goes in 2, 2 in 4, 3 in one, and finally 4 goes in 3.

It is easy to interpret this permutation in terms of the stocks in the figure 2. Noting 1=AXP, 2=HD, 3=WMT, 4=PG, the permutation {2,4,1,3} simply states that for a period of time from 5/15/2013 to 06/03/2013 price of the stock AXP permuted with the price of the stock HD, HD permuted with PG, WMT with AXP, and finally WMT change the place with PG.

The crossings of stocks appear in a different light now, it is easy to notice that are nothing else than permutations of stocks.

The next sections will explore the fascinating world of combinatorics and the ways that beautiful combinatorial objects are related to changings of stocks prices quotations in the market.

4. **"Decorating" the stock market permutations and some combinatorial remarks**

Wiring diagrams associated to stock permutations as the one in figure 2 have a beautiful visual impact. Still, in order to extract important combinatorial aspects related to stock arrangements, further mathematical tools are necessary.

The stocks permutation depicted in the wiring diagram in figure 2 written as {2,4,1,3} can be represented in a more simply way, named permutation diagram:

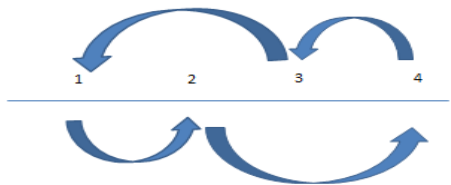

**Figure 3**. Permutation diagram for a selection of four DJIA stock components.

The permutation diagram above encodes the same information as wiring diagrams. Notice that the permutation graphic has two arcs pointed to the left and two arcs pointed to the right. The permutation diagram has the advantage that it can be extended to combinatorial objects, such as *decorated permutation* and *Grassmann necklace*, which will eventually leads to geometric representation of stock market.

**Decorated permutation** is a permutation $\pi = \begin{pmatrix} 12 \ldots n \\ \pi_1 \pi_2 \ldots \pi_n \end{pmatrix}$ with fixed points $\pi_i = i$ colored in two colors. This "dry" definition of a combinatorial object is actually hiding some common sense situations normally encountered if evaluate the components of a market index at different periods.

In the market evolution of stock prices there are stocks that do not cross with the other nearest stocks. Not having a transposition, for these stocks the points in the permutation are fixed, so that $\pi_i = i$. Let's take a look again at the figure 1 and analyze the permutation that result indexing the stocks quotations at 05.29.2013. The associated permutation, in this case, is {1,3,4,2}. It can be noticed that stock 1 (AXP) do not cross with its consecutive neighbor (HD) and remain at its initial place, in other words AXP is a fixed point, and $\pi_1 = 1$.

In the same register, resuming the market quotations at 06.05.2013 the associated permutation becomes {1,3,2,4}. This time, both stock 1 (AXP) and stock 4 (PG) are fixed points. These two fixed points must have cords associated in the permutation diagram and for that the following convention will take the place of the coloring function in the decorated permutation:

- the cord associated to a stock having the price increased from the initial moment is pointed to the right;
- the cord associated to a stock having the price decreasing from the initial moment is pointed to the left.

In the permutation {1,3,2,4} there are two fixed points, 1 and 4 associated to stocks

AXP and PG. The price of stock AXP increase from 72,78 at 05.15.2013 to 74,76 at 06/05/2013 and its cord points to the right; the price of PG decrease in the same period from 80,68 to 76,66 and the cord associated to it will points to the left.

The permutation diagram associated to permutation {1,3,2,4} is depicted in figure 4 and is describe a decorated permutation.

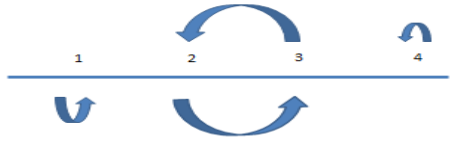

**Figure 4**. The decorated permutation {1,3,2,4}

Decorated permutation is in bijection with another remarkable combinatorial object, Grassmann necklace. The Grassmann necklace makes the connection with geometry that will be induced to stock market. As before, the combinatorial definition will be provided and the relation with stock market will be revealed.

A **Grassmann necklace** is a sequence $I = (I_1, \ldots, I_n)$ of subsets $I_i \subseteq \{1, \ldots, n\}$ such that

- if $i \in I_i$ than $I_{i+1} = I_i \backslash \{i\} \cup \{j\}$ for $j \subseteq \{1, \ldots, n\}$;
- if $i \notin I_i$ than $I_{i+1} = I_i$.

It will be soon noticed that it is easy to find the Grassmann necklace from a decorated permutation despite the raw combinatorial definition above. Every term $I_n$ in the Grassmann necklace sequence $I$ is formed by the cords pointing to the left in every cyclically shifted ordering of permutation terms.

The Grassmann necklace concept is depicted in figure 5 where every term in the cyclically shifted order is computed as being a part of the set $I$. The decorated permutation {2,4,1,3} is the same as before, and is based on the stock market permutation associated to the wiring diagram in fig 2.

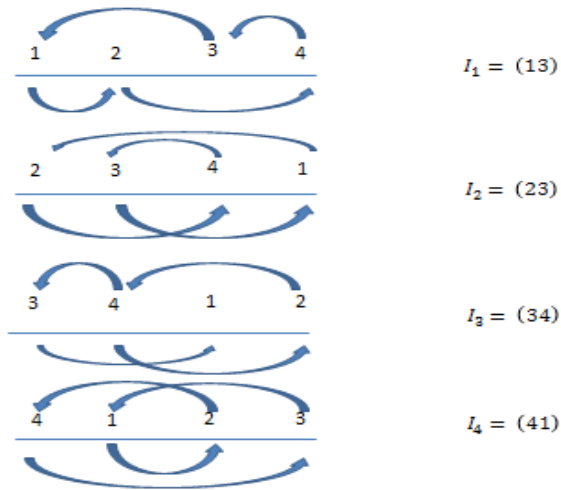

**Figure 5**. The terms of the Grassmann necklace associated to permutation {2,4,1,3}

It is easy to notice that are two cords pointing to the left and that will make every term in the Grassmann necklace to have two components. The first term

Just for exemplification consider the Grassmann necklace terms as being expressed by the stocks:

$$I_1 = (AXP\ WMT), I_2 = (HD\ WMT), I_3 = (WMT\ PG), I_4 = (PG\ AXP). \qquad (1)$$

Another way to construct the Grassmann necklace is explored in the next section by means of hook diagrams which constitutes an even richer object in aspects related to combinatorial concepts.

5. **Hook diagrams for stock permutations**

Hooks diagram not only help to find Grassmann necklace of a decorated permutation, as was pointed out in the last section, but also, more important, will provide information on the number of dimensions the final stock market geometrical object will have. The geometry of the stock market for the chosen four stock components of DJIA will be 3-dimensional. Considering all 30 stock components the geometry will be far more complicated, and the resulting geometrical object will live in a space with many more dimensions.

To construct the hooks diagram an alternative way to describe the decorated permutation is necessary. The decorated permutations are viewed this time as a map $\pi: \{1, \ldots, n\} \to \{1, \ldots, 2n\}$ such that $i \leq \pi_i \leq i + n$. Simply speaking the cords pointing to the left in a decorated permutation are sending beyond , that means that $\pi$ must be shifted by $n$ relative to the initial permutation.

Turning to the example of the DJIA four stocks components, the permutation {2,4,1,3} associated to the stock market must be shifted to {2,4,5,7}. Saying all this and skipping, for the sake of simplicity, the mathematical definitions, hook diagram associated to the above permutation can be depicted graphically as:

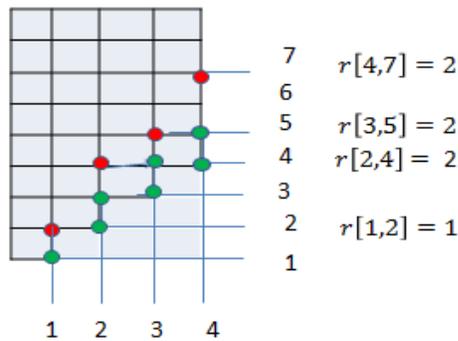

**Figure 6**. Hook diagram associated to permutation {2,4,1,3}.

Interested reader in construction of hooks out of decorated permutation associated to it, is referred to [1].

The hook diagram encodes the important information about the dimensionality of the geometrical object that stock market is associated with through the decorated permutation. In the diagram above the $r[i, \pi_i]$ is the number of other hooks which intersect the vertical (or horizontal) part of any hook $i \to \pi_i$, and are represented with green dots.

The dimension of the stock market geometric object is simply:

$$\dim(M) = \sum_{i=1}^{n} r[i, \pi_i] - k^2 \tag{2}$$

where $k$ is the number of cords in the decorated permutation pointing to the left.

To exemplify, for the permutation {2,4,5,7} the number of hook intersections are computed in the figure 6. Noticing that $k = 2$ this configuration of stock market has the dimension:

$$dim(M) = 7 - 4 = 3 \tag{3}$$

In other words it expected that the geometrical object that characterizes the stock market configuration defined by the permutation {2,4,5,7} to have 3 dimensions and could be visualize intuitively. It should be notice here that this is a particular case; generally the geometrical object of the stock market lives in higher dimensions and is impossible to be drawn directly.

It was stated in the latter sections that hook diagram is an alternative way of finding the Grassmann necklace for a certain stock market configuration. In terms of hooks every term $I_i$ in the Grassmann necklace $I$ is simply a list of $k$ horizontal lines which pass above the $i^{th}$ column. In the example of the permutation {2,4,5,7} the Grassmann necklace terms are shown in the figure 7 below:

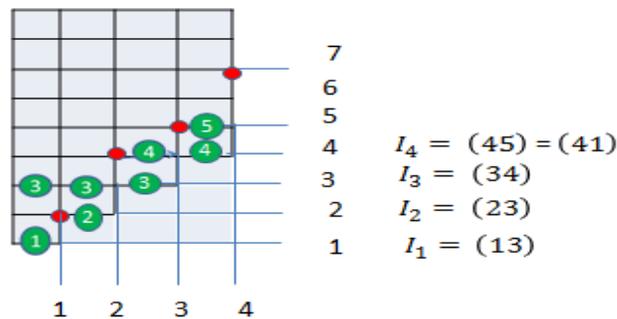

Figure 7. Grassmann necklace resulting from hook diagram.

Notice that that the same Grassmann necklace terms from the last section are recovered.

Grassmann necklace is important in the present exposition mainly because it effectively makes the connection to the beautiful structures in combinatorial geometry, such as *matroid*, *positroid* and *polytopes*, which will be explored in the next sections.

## 6. Combinatorial geometry and stock market polytope

The following section represents mainly a brief introduction to usual combinatorial and algebraic geometry concepts that will be further adapted to stock market realities. The interested reader is referred for extensive exploration of these beautiful branches of mathematics to []… A better understanding of …

The remarkable combinatorial objects that will be discussed further are the last bricks in the construction of the bridge relating stock market to geometry. The central figures in this section will be matroids ,positroid and eventualy polytopes; their relation with the Grassmann necklace will be also explored.

A **matroid** of rank k on the set $\{1, \dots, n\}$ is a nonempty collection $M \subseteq \binom{n}{k}$ of k elements subsets, called *bases* of $M$, that satisfies the exchange axiom; for any $I, J \in M$ and $i \in I$, there exist $j \in I$ such that $I \setminus \{i\} \cup \{j\} \in M$.

There is a reference here to the cyclically shifted order that take the mind to Grassmann necklace and a relation that could exist between these combinatorial objects. In fact there is a bijection between Grassmann necklace a special type of matroid, namely the *positroid*.

Positroid, a relatively new combinatorial concept and is mainly the fruit of A. Postnikov work and became "famous "due to its beautiful applications in Quantum Field Theory. The quintessence of Postnikov findings is that each decorated permutation and hence a Grassmann necklace corresponds to a positroid. So to speak every stock market representation, decorated with a permutation of stocks corresponds to a positroid.

A matroid $M \subseteq \binom{n}{k}$ is a **positroid** if and only if it can be written as $SM_{I_1}^1 \cap \dots \cap SM_{I_n}^n$ for a Grassmann necklace $I = (I_1, \dots, I_n)$; here $SM_{I_n}^n$ are the cyclically shifted Schubert cells.

Although this is the consecrated definition of positroid in the following, a derivation of it will be used, not only for simplicity, but also for the direct connection with Grassmann necklace.

$M$ is a positroid if and only ifthe following holds : $H \in M$ if and only if $H \geq_i I_i$ for all $i \in \{1, \dots, n\}$. In other words the positroid can be written directly from Grassmann necklace by taking:

$$M = \{H | H \geq_1 I_1, \dots, H \geq_n I_n\}$$

It is clear now why Grassmann necklace is so important.

Turning to the stock market example it will be easy to find the *stock market positroid* out of the decorated permutation associated with the crossing of stocks. Recall that the decorated permutation {2,4,1,3} related to state of the stock market has the Grassmann necklace:

$$I_1 = (13), I_1 = (14), I_1 = (23), I_1 = (34) \tag{4}$$

According to the definition, the stock market positroid will have the bases:

$$M = \{(13), (14), (23), (24), (34)\} \tag{5}$$

These abstract "coordinates" reflects the situation of stock market at 6/3/2013, and the stock market positroid encode the crossings of stocks up until this date.

One more step remains until the crossing of stock prices can be reflected in a geometrical object, which as will become immediately clear is a *positroid polytope* or the **stock market polytope**.

The positroid polytope is nothing else than the geometric representation of the positroid having the vertices defined by the positroid bases. In the case of the stock market positroid that has the bases found in (5), the associated **stock market polytope** is graphicaly depicted in figure 8:

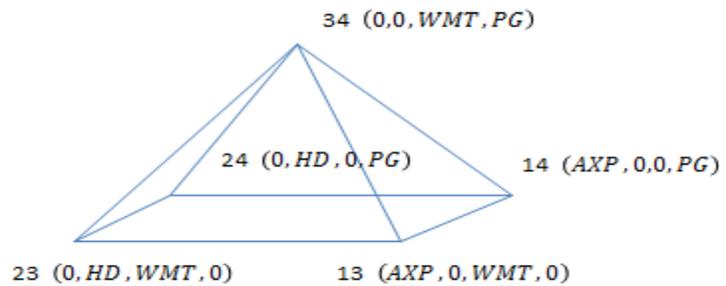

**Figure 8**. *The stock market polytope* (**stockMarkethedron**).

Notice that the polytope vertices are labeled either with the stocks that are port in the initial stock market permutation.

It is hard to believe that the geometrical construction above could represent the state of the stock market at some specific date, but still the **stock market polytope** encodes all the relevant information related to stocks price moving.

## 7. The Positive Grassmannian and the decomposition of the stock market polytope

An alternative and more fruitful way to construct the stock market polytopes is to consider the matrix approach of Grassmannians variety. The present paper intentionally avoids this approach due to its complexity and more deeply implications that will constitute the subject of a future exposition. The richness of the matrix in approaching the stock market will reside eventually in expressing the probability of finding the market in a certain state only by computing the volume of the stock market polytope.

Here the discussion is limited to showing the relation between the stock market polytope and the remarkable structures of *positive Grassmanians* that leads to stratification and decomposition into positroid cells. Simply speaking the decomposition of stock market polytope is nothing else but the elimination of stock crossings one of a time. The inverse operation, namely amalgamation is related to construction of the stock market polytope by adding crossing of stocks to the initial configuration of the market.

The Grassmannian $G(k, n)$ is the space of $k$-dimensional planes in $n$-dimensional space. The positive Grassmannian $G^+(k, n)$ is a subset of the real Grassmannian $G(k, n)$. In [1], [2] Postnikov shows that a positroid is a point in the *positive Grassmannian*, indexed by the Grassmann necklace. Also a positroid is associated to a *positive Grassmann cell* such as the $G^+(k, n)$ is the disjoint union of its cells.

It looks complex but as it will be seen immediately for the stock market example decorated by the permutation {2,4,1,3} that stock market polytope is simply the union of all the stock crossings. Removing the crossings one of a time reside in decomposition of the positive Grassmannian.

In order to better visualize the details of the decomposition, the permutation diagram of the stocks is used once again. This time the permutation diagram is associated to another combinatorial object, the non-crossing partition in the way shown in figure 9.

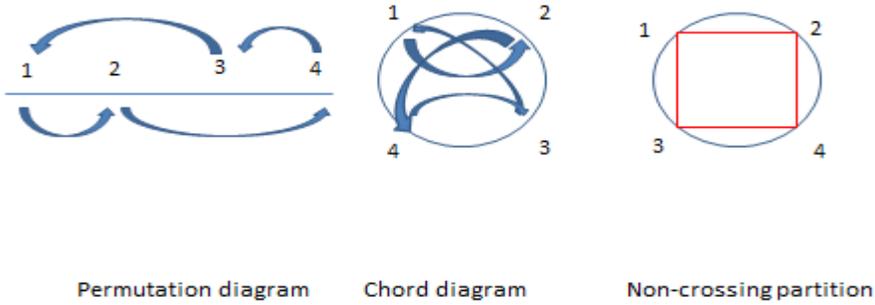

Permutation diagram    Chord diagram    Non-crossing partition

**Figure 9**. Non-crossing partition of the stocks decorated permutation.

The particularity of non-crossing partition is that its edges do not intersect. This is exactly what happens if a cross of stocks is removed from the initial permutation.

In the non-crossing partition in figure 9 removing the edge (34) it follows that the edge (24) vanish too, otherwise the non-crossing partition no longer exist. The situation is explicit in figure 10.

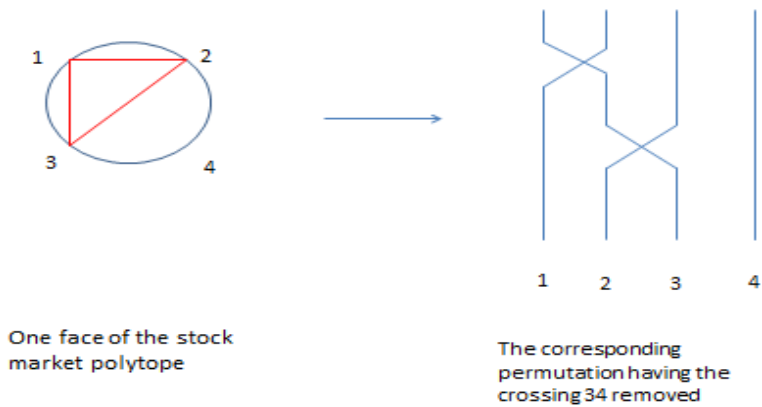

One face of the stock market polytope

The corresponding permutation having the crossing 34 removed

**Figure 10**. Removal of stock crossing associated with a face of stock market polytope.

It is clear now from the wiring diagram that removing (34) cancels the (24) in the same time. This cancelation in pairs preserves the positivity of the Grassmannian $G^+(2,4)$ after each crossing removal. Each face of a positroid is a positroid.

Removals of crossings in the permutation associated to stock market reside in the decomposition of the positive Grassmannian $G^+(2,4)$ labeled by the stock market polytope in positroid cells as is depicted in the figure 11.

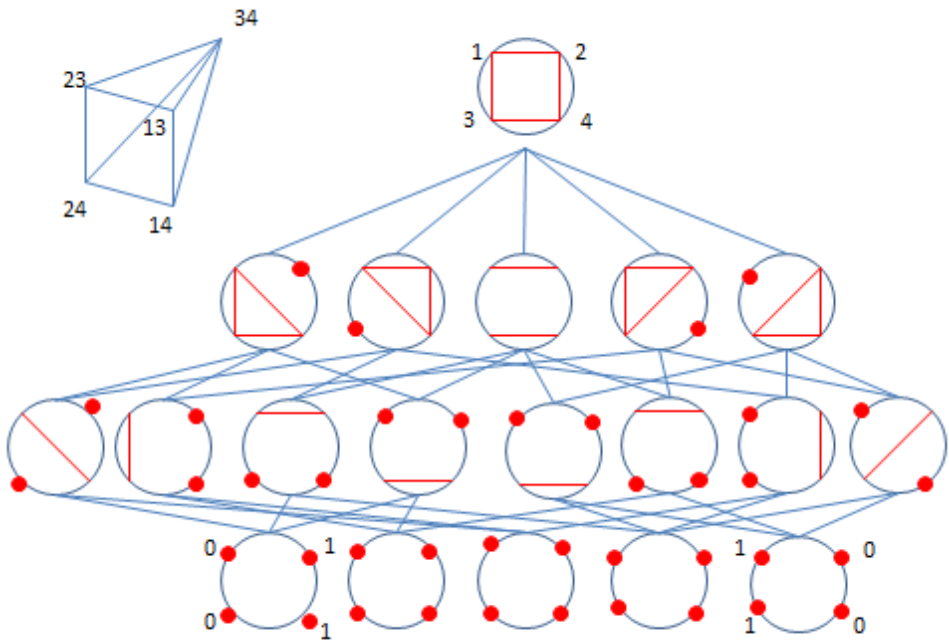

**Figure 11**. Decomposition of the stock market polytope

Notice that every crossing corresponds with a face of the polytope. The stock market polytope can be constructed step-by-step from the initial state of the market by adding edges and faces every time a market quotation induces a stock crossing. Adding positroid cells is gluing procedure

Decomposition and the gluing procedure are more complex when all the DJIA stock components are considered. An increasing number of stock crossings have to be taken into account, a procedure that leads to a space having higher dimensions.

## 8. Discussions and concluding remarks

The present paper constitutes a map connecting the price evolution of stocks in the market with complex and beautiful geometrical shapes. The "roads" of this map are combinatorial concepts as permutations and hooks, positroid and polytopes or the remarkable geometric structures of positive Grassmannians.

The key point in constructing the above mentioned map is to associate changings in stocks prices with combinatorial objects such as permutations. This association is easy to make if the stocks that compose an index are arranged in a particular way and take into account the crossing of stocks. Under some conventions the diagram that results by winding the prices of stocks is exactly a wiring diagram associated to permutations.

To exemplify with real stock quotations, *Dow Jones Industrial Average* (DJIA) index is analyzed and prices of four of its stock components (APX, HD, WMT and PG) for a short period of time are integrated in a wiring diagram that is proved to be a remarkable permutation diagram. In other words it can be said about the stocks in discussion that AXP permutes with HD, or that WMT permutes with PG.

Pushing further the combinatorial approach of the stock market, the resulting wiring diagram is transformed according to some simple rules in a permutation diagram, with the help of decorated permutation. Some combinatorial objects related to *decorated permutation* are briefly explained, along with their connections with stock market, without diving too deep into the associated mathematics.

Combinatorial notions, like Grassmann necklace, matroid or positroid are look exotic in the financial frame, but as it was seen in the present exposition they fit well in the picture of stock market.

Finally the map connecting the price quotations of stocks in the market with beautiful geometrical shapes is completed by adding the concept of polytopes. This way the geometrical object that encodes all the relevant information of stock market can be named as ***stock market polytope***.

To better understand the polytopes, the *positive Grassmannians* are added to complete the picture in which the performances of stock market can be viewed by geometrical shapes. Since typically polytopes live in higher dimensions their shapes cannot have a direct image. The positive Grassmannian cells help in the construction of multidimensional stock market polytopes by simply gluing together all the stock crossings between components of the market index.

Certainly is not easy to imagine the stock market as a geometric shape, such as a pyramid to name a trivial example. Still the stock market polytope fully encodes all the relevant information about the current (or future as a future paper will assess) state of the stock market.

Deeper insights into the properties of stock market polytopes such as the relation between its volume and the probability to a certain state of market to occur are exploring in a future paper.